# Observation of integer and fractional quantum anomalous Hall effects in twisted bilayer MoTe$_2$

Fan Xu[1,2], Zheng Sun[1], Tongtong Jia[1], Chang Liu[1], Cheng Xu[3,4], Chushan Li[1], Yu Gu[1], Kenji Watanabe[5], Takashi Taniguchi[6], Bingbing Tong[7], Jinfeng Jia[1,2,8], Zhiwen Shi[1,2], Shengwei Jiang[1,2], Yang Zhang[3,9]*, Xiaoxue Liu[1,2,8]*, Tingxin Li[1,2,8]*

[1]Key Laboratory of Artificial Structures and Quantum Control (Ministry of Education), Shenyang National Laboratory for Materials Science, School of Physics and Astronomy, Shanghai Jiao Tong University, Shanghai 200240, China.
[2]Tsung-Dao Lee Institute, Shanghai Jiao Tong University, Shanghai, 201210, China.
[3]Department of Physics and Astronomy, University of Tennessee, Knoxville, TN 37996, USA
[4]State Key Laboratory of Low Dimensional Quantum Physics and Department of Physics, Tsinghua University, 100084 Beijing, China
[5]Research Center for Electronic and Optical Materials, National Institute for Materials Science, 1-1 Namiki, Tsukuba 305-0044, Japan
[6]Research Center for Materials Nanoarchitectonics, National Institute for Materials Science, 1-1 Namiki, Tsukuba 305-0044, Japan
[7]Beijing National Laboratory for Condensed Matter Physics and Institute of Physics, Chinese Academy of Sciences, Beijing 100190, China
[8]Hefei National Laboratory, Hefei 230088, China
[9]Min H. Kao Department of Electrical Engineering and Computer Science, University of Tennessee, Knoxville, Tennessee 37996, USA

*Emails: txli89@sjtu.edu.cn; xxliu90@sjtu.edu.cn; yangzhang@utk.edu

**Abstract**

**The interplay between strong correlations and topology can lead to the emergence of intriguing quantum states of matter. One well-known example is the fractional quantum Hall effect, where exotic electron fluids with fractionally charged excitations form in partially filled Landau levels. The emergence of topological moiré flat bands provides exciting opportunities to realize the lattice analogs of both the integer and fractional quantum Hall effects without the need for an external magnetic field. These effects are known as the integer and fractional quantum anomalous Hall (IQAH and FQAH) effects. Here, we present direct transport evidence of the existence of both IQAH and FQAH effects in small-angle-twisted bilayer MoTe$_2$. At zero magnetic field, we observe well-quantized Hall resistance of $h/e^2$ around moiré filling factor $v$ = -1 (corresponding to one hole per moiré unit cell), and nearly-quantized Hall resistance of $3h/2e^2$ around $v$ = -2/3, respectively. Concomitantly, the longitudinal resistance exhibits distinct minima around $v$ = -1 and -2/3. The application of an electric field induces topological quantum phase transition from the IQAH state to a charge transfer insulator at $v$ = -1, and from the FQAH state to a topologically trivial correlated insulator, further transitioning to a metallic state at $v$ = -2/3. Our study paves the way for the investigation of fractionally charged excitations and anyonic statistics at zero magnetic field based on semiconductor moiré materials.**

**Introduction**

IQAH effect [1-3], characterized by well-quantized Hall resistance of $h/ne^2$ (where $n$ is an integer, $h$ and $e$ denoting the Planck's constant and electron charge, respectively) and vanishing longitudinal resistance at zero magnetic field, have already been successfully demonstrated in several material systems [4-7]. In conjunction with strong electron-electron interactions, the partially filled Chern bands can host the FQAH states [8-12], in analog to the fractional quantum Hall effects from partially filled Landau levels [13, 14]. However, realizing FQAH effect turns out to be considerably more challenging in real materials due to the requirement of nearly uniform Berry curvature distributions within a topological flat band. Chern bands formed in graphene moiré superlattices have been considered [15-21] as a tunable platform to realize FQAH effect. A recent experiment [22] has reported fractional Chern insulator (FCI) states in twisted bilayer graphene, but only under a 5 T external magnetic field. This external magnetic field is essential for redistributing the Berry curvature inside the topological moiré band to enable the FCI states. On the other hand, semiconductor moiré materials [23,24] offer another promising avenue for engineering topological and strong correlation physics. Experimental studies have demonstrated [7,25] tunable band topology and IQAH effect in near-60-degree-twisted (commonly referred to as AB-stacked) $MoTe_2/WSe_2$ moiré heterostructures. Theoretical works [26-29] also suggest semiconductor moiré systems have the potential to host both IQAH and FQAH effects. Specifically, the low energy physics in small-angle-twisted (commonly referred to as AA-stacked) bilayer $MoTe_2$ and $WSe_2$ is proposed to be described by the Kane-Mele-Hubbard model, and exhibits flat Chern bands at small twist angles. Recently, electrically tunable magnetism has been reported in twisted bilayer $MoTe_2$ ($tMoTe_2$) [30]. Moreover, signatures of IQAH and FQAH states have been observed through photoluminescence [31] and optical compressibility measurements [32] in $tMoTe_2$, attracting great interest [33-37]. Another experiment [38] reported integer Chern insulators at zero magnetic field in twisted bilayer $WSe_2$ through local compressibility measurements. However, definitive and indisputable evidence of both IQAH and FQAH effects in these systems, namely the quantized Hall conductance, is still lacking.

In this study, we report transport evidence of the existence of both IQAH and FQAH effects in $tMoTe_2$. We fabricated AA-stacked $MoTe_2$ bilayers devices with twisted angle of ~ 3.7-3.85 degrees and conducted systematic transport measurements. The schematic structure of the device is depicted in Fig. 1a, where the moiré filling factor ($v$) and the vertical electric displacement field ($D$) can be independently controlled by the top gate voltage $V_t$ and back gate voltage $V_b$. To achieve Ohmic contacts, the $MoTe_2$ layer is directly touched with few-layer $TaSe_2$ and we employed the global $Si/SiO_2$ gate to induce heavily hole doping in the contact region (see Appendix and Supplemental Fig. 1 for detailed information of device fabrications and transport measurements). At $v = -1$, we observe quantized Hall conductance plateau (within certain $v$ and $D$ ranges) of $e^2/h$ at zero magnetic field. Similarly, at $v = -2/3$, quantized Hall conductance plateau of $2e^2/3h$ has been observed in the zero magnetic field limit, albeit within a narrower range of $v$ and $D$ as compared to $v = -1$. The longitudinal resistance concomitantly exhibits distinct minima within similar $v$ and $D$ ranges. Moreover, both the IQAH and FQAH states can be tuned into topologically trivial states by applied $D$-field.

**IQAH effect at *v* = -1 and FQAH effect at *v* = -2/3**

Figure 1b and 1c present the longitudinal sheet resistance ($\rho_{xx}$) and the Hall resistance ($\rho_{xy}$) of device I, as a function of $v$ and $D$ in the zero magnetic field limit. The moiré density ($n_M$) of device I is estimated to be approximately $3.8 \times 10^{12}$ cm$^{-2}$ (see Appendix), corresponding to a twisted angle about 3.7 degrees. At large $D$ values, the $t$MoTe$_2$ experiences a substantial interlayer potential difference, resulting in a layer-polarized state. In this regime, the physics is predominantly governed by a single-band Hubbard model on triangular lattices [39]. Consequently, we observe a series of topologically trivial correlated insulating states at $v$ = -1, $v$ = -2/3, -1/2 and -1/3. These results are qualitatively consistent with previous observations in transition metal dichalcogenide (TMDc) moiré heterostructures with significant band offset [40-47]. Notably, as the $D$ value decreases, substantial $\rho_{xy}$ emerge within the $v$ range of ~ -1.25 to ~ -0.5, particularly near $v$ = -1 and $v$ = -2/3, demonstrating spontaneous time reversal symmetry (TRS) breaking in these regions. It's evident that the critical electric field ($D_c$) required to suppress the spontaneously TRS breaking phase gradually decreases with decreasing hole density, which is consist with recent optical studies [30-32]. Additionally, within the region where TRS is spontaneously broken, $R_{xx}$ exhibits minima around $v$ = -1 and $v$ = -2/3. These are distinctive attributes of chiral edge transport—a significant indicator of the presence of quantum anomalous Hall effects.

We further investigate the spontaneously TRS breaking states around $v$ = -1 and $v$ = -2/3. Figure 2a and 2b illustrate $\rho_{xy}$ and $\rho_{xx}$, respectively, as a function of out-of-plane magnetic field ($B$) at $v$ = -1 and $D$ = -115 mV/nm (see Supplemental Fig. 2a-2c for other $D$ values). A clear magnetic hysteresis loop with coercive field around 30 mT at 30 mK is observed. At zero magnetic field, $\rho_{xy}$ remains quantized at $h/e^2$ (±1.5%) below 2 K. Correspondingly, $\rho_{xx}$ is almost vanishing at low temperatures (< 100 Ω below 900 mK), and show two sharp peaks at the coercive fields. With increasing temperature, the $\rho_{xy}$ at $B$ = 0 begins to deviate from its quantized value, while simultaneously, $\rho_{xx}$ at $B$ = 0 undergoes rapid growth. The Curie temperature at $v$ = -1 is determined to be about 10-12 K by the temperature-dependent $\rho_{xy}$ at $B$ = 0 (see Supplemental Fig. 2d).

We now turn to the $v$ = -2/3 state. Figure 2c and 2d show $\rho_{xy}$ and $\rho_{xx}$, respectively, at $v$ = -2/3 and $D$ = 5 mV/nm, as a function of $B$ at temperature ($T$) ranging from 600 mK to 6 K. Below 600 mK, electrical contacts at such low fillings become non-Ohmic in this device (see Supplemental Fig. 1c), which disturbing reliable transport measurements. Again, at low temperatures (< 1 K), we observe clear magnetic hysteresis with coercive field ~5 mT and nearly $B$ independent $\rho_{xy}$ except at the magnetic switching. At 600 mK, the $\rho_{xy}$ at zero magnetic field is close to a quantized value of $3h/2e^2$. At elevated temperatures, $\rho_{xy}$ starts to deviate from the quantized value, and a Curie temperature of ~4 K is determined from the $T$-dependent $\rho_{xy}$ data. On the other hand, although a distinct local minimum of $\rho_{xx}$ can be observed around $v$ = -2/3, we notice it remains substantial even at 600 mK, approximately around 20 kΩ at $B$ = 0. And the $\rho_{xy}$ quantization around $v$ = -2/3 is less accurate compared to the $v$ = -1 state (see Supplemental Fig. 3 for $\rho_{xx}$ and $\rho_{xy}$ as a function of $v$ at varying temperatures, and Supplemental Fig. 4 for longitudinal resistance measured in different configurations).

These observations demonstrate that the state at $v$ = -1 is an IQAH state with Chern number $|C|$ = 1 and the state at $v$ = -2/3 is a FQAH state with $|C|$ = 2/3. The imperfection quantization of $\rho_{xy}$ and non-negligible $\rho_{xx}$ of the $v$ = -2/3 FQAH state arises presumably from the disorder and inhomogeneity of the sample that leads to remnant dissipation in the bulk. It may also relate to the relatively large contact resistance at $v$ = -2/3 at low temperatures. The $v$ = -1 state is more robust because of a much larger gap compared to the $v$ = -2/3 state. Independently, the Chern number can be deduced from the Streda formula, $n_M \frac{dv}{dB} = C \frac{e}{h}$, which has been commonly used to determine the Chern number of quantum anomalous Hall states or Chern states in moiré systems [5,7,21,22,31,32,38]. We checked the dispersion of $\rho_{xx}$ in $v$ and $B$, and extract the Chern number for the $v$ = -1 and $v$ = -2/3 states to be $|C|$ = 1.05 ± 0.09 and $|C|$ = 0.6 ± 0.05, respectively (see Supplemental Fig. 6). The results are reasonably consistent with the observed quantized Hall resistance, and recent optical studies [31,32]. The difference between the experimental values and the anticipated values ($|C|$ = 1 and 2/3) may also be related to sample inhomogeneities and disorder effects. These results are qualitatively reproduced in another pair of contacts (Supplemental Fig. 5, 6), and also in another device (Supplemental Fig. 7).

**Quantized Hall conductance plateau**
We further characterize the Hall conductance $\sigma_{xy}$ against $v$ and $D$. Figure 2e shows $\sigma_{xy}$ as a funciton of $v$ at $D$ = 5 mV/nm, across varying temperatures. We derive $\sigma_{xy}$ using the reciprocal resistance-to-conductance tensor conversion given by $\sigma_{xy} = \frac{\rho_{xy}}{\rho_{xx}^2+\rho_{xy}^2}$, where the $\rho_{xy}$ ($\rho_{xx}$) is antisymmetrized (symmetrized) at $B$ = ± 0.3 T (see Appendix and Supplementary Fig. 3). Remarkably, $\sigma_{xy}$ exhibits quantization plateau at $e^2/h$ (±2%) approximately within the range of $v$ = -1.1 to -0.95, and this quantized range narrows with increasing temperature. Around the $v$ = -2/3 filling, a narrower $\sigma_{xy}$ plateau is evident, spanning from roughly $v$ = -0.65 to -0.7, where the plateau value closely aligns with $2e^2/3h$ (±5%). Figure 2f further displays $\sigma_{xy}$ as a function of $D$ at $v$ = -1 and $v$ = -2/3 in the zero magnetic field limit at 900 mK. The $\sigma_{xy}$ remains at the quantized values within a specific range of $D$, until the $D$-field induced topological phase transition takes place. The critical electric field $D_c$ of the topological phase transition is estimated to be about 120 mV/nm and 15 mV/nm for $v$ = -1 and $v$ = -2/3, respectively. These results illustrate that the Hall quantization of $v$ = -1 and $v$ = -2/3 is robust against the perturbation of $v$ and $D$.

Besides the $v$ = -1 and $v$ = -2/3 quantization plateaus, large anomalous Hall effects appear at wide parameter regions, which is consistent with strong ferromagnetism observed from the $\rho_{xy}$ map (Fig. 1c). At $v$ < -1, we observed an increasing anomalous Hall conductance significantly larger than $\frac{e^2}{h}$, which indicates that the second moiré band has the same Chern number (instead of opposite Chern number) as the first moiré band. The overall trend can be captured by mean field calculation with the projected interaction model (see Appendix and Supplemental Fig. 8). Here the anomalous Hall metal phases preserve translation symmetry, and is a fully polarized metal governed by strong direct exchange interaction between $B^{Mo/Te}$ and $B^{Te/Mo}$ Wannier orbitals. The larger than $\frac{e^2}{h}$ anomalous Hall conductance between -1 < $v$ < -0.8 and the evident temperature dependence in Fig. 2e implies that there is an additional extrinsic contribution from mobile charge carriers besides the intrinsic mechanism of Berry curvature.

***D*-field induced topological phase transition**

Next, we examine the topological phase transitions driven by *D*-field at both $v$ = -1 and $v$ = -2/3. Figure 3a and 3b display $\rho_{xx}$ maps versus *D* and *T* at $v$ = -1 and $v$ = -2/3, respectively. At $v$ = -1, the system hosts IQAH effect at $|D|$ < ~ 120 mV/nm, where $\rho_{xx}$ decrease rapidly with decreasing *T* below ~ 10 K (close to the Cuire *T* at $v$ = -1) due to dissipationless chiral edge transport. Above 10 K, $\rho_{xx}$ exhibits a weaker temperature dependence but still increases with raising *T* (see Supplemental Fig. 9). This behavior contrasts with the temperature dependence of the IQAH effect in AB-stacked MoTe$_2$/WSe$_2$ [7], where $\rho_{xx}$ exhibits an insulating behavior above the Cuire *T*. On the other hand, at larger *D* values, the system transitions into a topologically trivial insulator, resulting in a thermal activation behavior with $\rho_{xx}$ growing rapidly with decreasing *T*.

Regarding the $v$ = -2/3 states, five distinctive regions can be identified, as illustrated in Fig. 3b and 3d. At *D* close to zero, the system exhibits FQAH effect, and the application of *D* field drives the FQAH state to a topologically trivial correlated insulator state. Further increasing *D* field suppresses the trivial insulating state and finally drive the system into a metallic state. The FQAH state preserves $C_{2y}$ symmetry, and the applied vertical electric field will break the layer symmetry to form a fractional correlated trivial insulator (commonly referred to as generalized Wigner crystal or charge density wave insulator in literatures) with honeycomb lattice polarized at one layer [48]. Since the bandwidth of the first moiré band gets larger with the applied *D*-field (see Supplemental Fig. 8), the Wannier functions polarized at a single layer will become more spread out. Therefore, the correlated trivial insulator at $v$ = -2/3 transits to a metallic state at large *D*-field.

We extract the gap values in both the topologically trivial and nontrivial insulating phases at $v$ = -1 and $v$ = -2/3 by thermal activation fitting of *T*-dependent $\rho_{xx}$ (see Supplemental Fig. 9), as shown in Fig 3c and 3d. Near the boundary of the topological phase transition, the thermal activation fitting become invalid due to the coexistence of both bulk and edge transport. For $v$ = -1, the gap size ($\Delta$) of the IQAH phase at *D* = 0 is around 20 K, and it gradually decreases with the increasing of *D*. In the topologically trivial regime, the gap size rapidly increases, roughly follow a linear relationship with increasing *D*. This is fully consistent with the behavior of a charge transfer gap. At *D* = 0, $C_{2y}$ symmetry is preserved and charges are equally distributed at B$^{Mo/Te}$ and B$^{Te/Mo}$ moiré region. At finite *D* > 0, charges transfer from B$^{Mo/Te}$ to B$^{Te/Mo}$, which leads to a gap closing and reopening, forming a charge transfer insulator. The gap size of charge transfer insulator $\Delta_{CTI}$ will be linearly proportional to applied *D*- field, which is confirmed in our Hartree-Fock field simulations (see Supplemental Fig. 8). The gap size of $v$ = -2/3 states are significantly smaller compared to $v$ = -1, where the maximum $\Delta$ value of the FQAH effect and the fractional correlated trivial insulator is about 1 K and 5 K, respectively. Future experimental studies on higher quality samples, and theoretical investigations are required to understand the nature of this phase transition and related critical phenomena.

**Conclusions**

In conclusion, we have observed both IQAH and FQAH effects in AA-stacked *t*MoTe$_2$ through electrical transport measurements. We have further explored the electric-field-driven topological phase transitions involving IQAH, FQAH, and other, topologically trivial correlated states. Future studies are needed to confirm the charge fractionalization and statistic properties of the FQAH state. The present work demonstrates $t$MoTe$_2$ as a fertile ground for exploring exotic quantum phenomena arising from electronic correlations and topology.

**Additional Notes:** During the submission of this manuscript, we become aware of a recent work that also reports the observation of IQAH and FQAH effects in $t$MoTe$_2$ by electrical transport measurements [49].

**Acknowledgement**
We thank Prof. Rui-Rui Du, Prof. Kin Fai Mak, and Prof. Jie Shan for helpful discussions. This work is supported by the National Key R&D Program of China (Grant No. 2022YFA1405400, 2022YFA1402702, 2022YFA1402404, 2021YFA1401400, 2021YFA1400100, 2021YFA1202902, 2019YFA0308600), the National Natural Science Foundation of China (Grant No. 12174249, 92265102, 12174250, 12141404, 12074244), the Natural Science Foundation of Shanghai (Grant No. 22ZR1430900), the Innovation Program for Quantum Science and Technology (Grant No. 2021ZD0302600, 2021ZD0302500), and the start-up fund of Shanghai Jiao Tong University. X.L. acknowledges the Pujiang Talent Program (Grant No. 22PJ1406700). T.L. and S.J. acknowledges the Shanghai Jiao Tong University 2030 Initiative and Yangyang Development Fund. Y.Z. was supported by the start-up fund at University of Tennessee Knoxville, and the National Science Foundation Materials Research Science and Engineering Center program through the UT Knoxville Center for Advanced Materials and Manufacturing (DMR-2309083). K.W. and T.T. acknowledge support from the JSPS KAKENHI (Grant No. 21H05233 and 23H02052) and World Premier International Research Center Initiative (WPI), MEXT, Japan. A portion of this work was carried out at the Synergetic Extreme Condition User Facility (SECUF).

**APPENDIX: METHODS**

**1. Device fabrications**
We fabricated the $t$MoTe$_2$ devices by standard dry transfer method [50] with polycarbonate (PC) stamps. In brief, thin flakes of graphite (3-5 nm), hBN (15-30 nm), TaSe$_2$ (2-5 nm), and MoTe$_2$ (monolayer) were mechanically exfoliated onto Si/SiO$_2$ substrates and identified by optical microscope. The thickness of thin flakes is measured using an atomic-force microscope (AFM). The MoTe$_2$ monolayer was mechanically cut into two parts by AFM tips. The pick-up sequence is top graphite, top hBN, two pieces of TaSe$_2$, two pieces of monolayer MoTe$_2$, bottom hBN and bottom graphite. The twisted angle between the two-layer of MoTe$_2$ monolayer is controlled by a mechanic rotator. Then the entire stack was released onto a Si/SiO$_2$ substrate. We handled MoTe$_2$ flakes inside a nitrogen-filled glovebox with oxygen and water levels below one part per million (ppm) to minimize degradation of MoTe$_2$. Windows were prior opened on the top hBN by e-beam lithography (EBL) and reactive ion etching (RIE), then Ti/Au (5 nm/50 nm) electrodes were e-beam evaporated in order to contact the TaSe$_2$ layer. To define the Hall bar geometry of the twisted MoTe$_2$ channel and eliminate the unavoidable monolayer MoTe$_2$ region (which could potentially cause an electrical short to the moiré region), we carried out standard EBL and RIE processes. The Hall bar width ($W$) is nominal 1 μm, and the separation between voltage probes (L) is nominal 2 μm. Although the MoTe$_2$ layer is directly touched with metallic TaSe$_2$ layer, a large doping is still needed in the contact region to form good ohmic contact for MoTe$_2$. Therefore, the global Si/SiO$_2$ (285 nm) gate was used to induce heavily hole doping in the contact region during the measurements. The carrier density of the Hall bar channel is only controlled by the top gate and back gate, since the back-gate electrodes could screen the electric field from the Si/SiO$_2$ gate. With this device geometry, Ohmic contacts can be achieved irrespective of the applied electric fields. In contrast, the previously used geometry in TMDc heterostructure devices [7,25,51] can only achieve Ohmic contact for a restricted range of high electric fields.

## 2. Transport measurements

Electrical transport measurements were performed in a closed-cycle $^4$He cryostat (Oxford TeslatronPT, base temperature about 1.5 K) with superconducting magnet. Measurements below 1.5 K were performed in a top-loading dilution fridge (Oxford TLM, base temperature about 20 mK) with superconducting magnet. Standard low-frequency (<17 Hz) lock-in (SR830 and SR860) techniques were used to measure the sample resistance. The bias current is limited within 3 nA, especially for measurements below 1.5 K, to avoid sample heating, and avoid disturbing the fragile states. Voltage pre-amplifier with 100 MΩ impedance were used to measure sample resistance up to several MΩ. Finite longitudinal-transverse coupling occurs in our devices that mixes the longitudinal resistance $R_{xx}$ and Hall resistance $\rho_{xy}$. To correct this effect, we used the standard procedure to symmetrize $\left(\frac{R_{xx}(B)+R_{xx}(-B)}{2}\right)$ and antisymmetrize $\left(\frac{\rho_{xy}(B)-\rho_{xy}(-B)}{2}\right)$ the measured $R_{xx}$ and $\rho_{xy}$ under positive and negative magnetic fields to obtain accurate values of $R_{xx}$ and $\rho_{xy}$, respectively. Alternatively, Onsager symmetrization method could be performed by exchanging the voltage contacts with the current contacts (See Supplemental Fig. 4). The longitudinal sheet resistance is derived by $\rho_{xx} = R_{xx}\frac{W}{L}$.

## 3. Determination of moiré filling factors

In dual-gated devices, the carrier density $\left(n = \frac{c_t V_t + c_b V_b}{e}\right)$ and the effective vertical electric displacement field $\left(D = \frac{c_t V_t - c_b V_b}{2\varepsilon_0} - D_{built-in}\right)$ can be independently controlled by the top gate voltage $V_t$ and back gate voltage $V_b$. Here, $D_{built-in}$, $\varepsilon_0$, $c_t$ and $c_b$ denote the built-in electric field, vacuum permittivity, top gate capacitance and back gate capacitance, respectively. The built-in electric field is likely arising from the structure asymmetry. For Device I, the $D_{built-in}$ is about 110 mV/nm, and for device II, the $D_{built-in}$ is less than 10 mV/nm. The value of $c_t$ and $c_b$ are mainly determined by measuring hBN thickness. We convert $n$ to moiré filling factor $v$ using the density difference between a series of correlated insulating states with prominent $\rho_{xx}$ peaks.

## 4. Wannier projected tight-binding model and Hartree-Fock simulation

The existence of FQAH effect has been discussed with band projected exact diagonalization [34,35], here we mainly focused on the understanding of field and filling dependent transition at integer filling and other generic filling factors. We first project the continuum model for the twisted MoTe$_2$

(parameters in Ref. [34]) to maximally localized Wannier functions [48], which captures the band dispersion and topological feature of the topmost moiré valence band. The tight-binding model is then constructed as:

$$H = t_1 \sum_{\langle i,j \rangle, \alpha} \hat{c}^\dagger_{i\alpha} \hat{c}_{j\bar{\alpha}} + t_2 \sum_{\langle\langle i,j \rangle\rangle, \alpha} e^{(-1)^\alpha i v_{ij} \phi} \hat{c}^\dagger_{i\alpha} \hat{c}_{j\alpha}$$

where α is the index of the sublattice (-1 for $B^{Mo/Te}$ region, +1 for $B^{Te/Mo}$ region); $i, j$ is the unit cell index; $t_1$ is the nearest neighbor hopping parameter, $t_2$ and $\phi$ is the amplitude and the phase of second nearest neighbor hopping. This realizes a generalized Kane-Mele model. Since the Wannier functions at $B^{Mo/Te}$ and $B^{Te/Mo}$ are 90% layer polarized, the gating field $D$ will introduce a site potential difference with strength linear in $D$. And the interacting terms in real space are calculated as

$$V = \frac{1}{2} \iint dr dr' \frac{e^2}{\epsilon|\mathbf{r} - \mathbf{r}'|} \hat{c}^\dagger_\sigma(\mathbf{r}) \hat{c}^\dagger_{\sigma'}(\mathbf{r}') c_{\sigma'}(\mathbf{r}') c_\sigma(\mathbf{r})$$

We note the ferromagnetic exchange interaction term $H_{ex} = \sum_{m,n} \sum_{\sigma\sigma'} V_{mn} a^\dagger_{m\sigma} a^\dagger_{n\sigma'} a_{m\sigma'} a_{n\sigma}$ is crucial for strong ferromagnetism at the wide range of filling factors. To capture the increasing anomalous Hall conductance at $|\nu| > 1$, we employ the three-band tight-binding model with Chern numbers -1, -1, and 2. With the interacting Hamiltonian in real space, we then performed the Hartree-Fock simulation under parameter space of displacement field and filling factor (which modifies the model parameter) using different supercell configurations. The simulation results are presented in Supplemental Fig. 8, which qualitatively capture the filling and displacement field-induced transition.

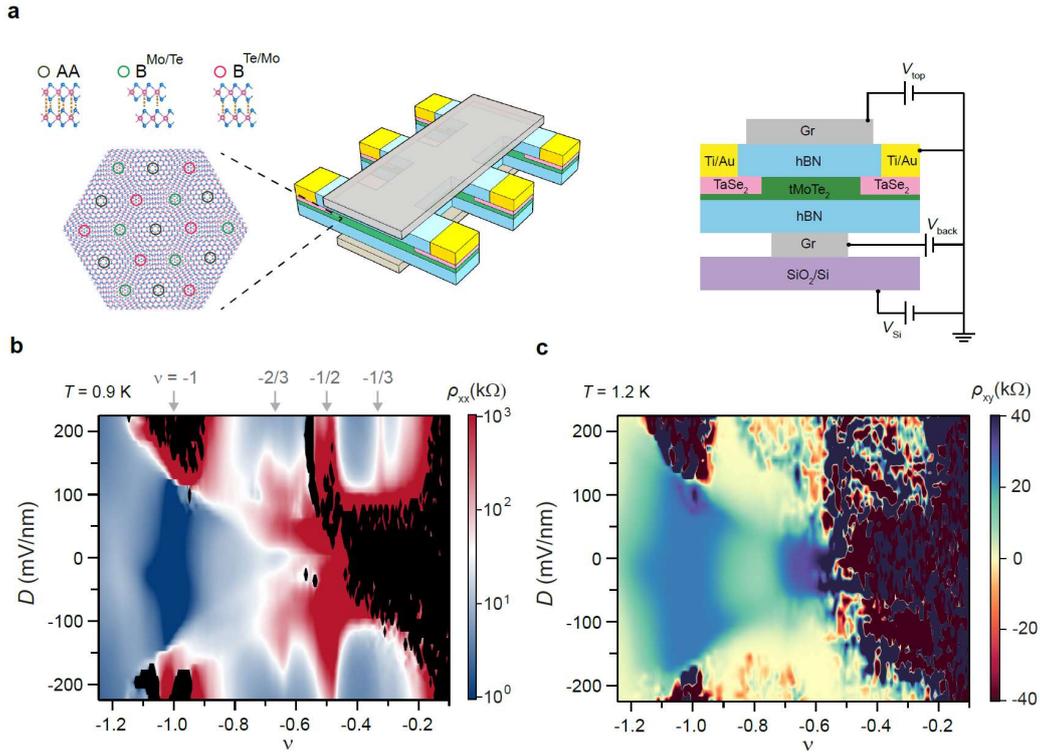

**Figure 1 | Phase diagram of AA-stacked *t*MoTe₂. a,** Left: Schematic of the Hall bar device used for transport measurements. Right: Side view of the device structure. The carrier density of the Hall bar channel is only controlled by the top gate and the back gate, whereas the carrier density of TaSe$_2$/MoTe$_2$ contact regions and Hall bar arms are mainly controlled by the global Si/SiO$_2$ gate. The inset shows the schematic moiré superlattices of *t*MoTe$_2$. High symmetry stackings are highlighted by circles. **b,** Longitudinal sheet resistance and **c,** Hall resistance as a function of $v$ and $D$. $\rho_{xx}$ is measured under zero magnetic field at 900 mK; $\rho_{xy}$ is the antisymmetrized results under an out-of-plane magnetic field of $\pm 0.1$ T at 1.2 K. The arrows mark $v$ = -1, -2/3, -1/2, and -1/3, where correlated insulating states (topologically trivial) formed at relatively large *D*-field. The black regions are very insulating and experimentally inaccessible.

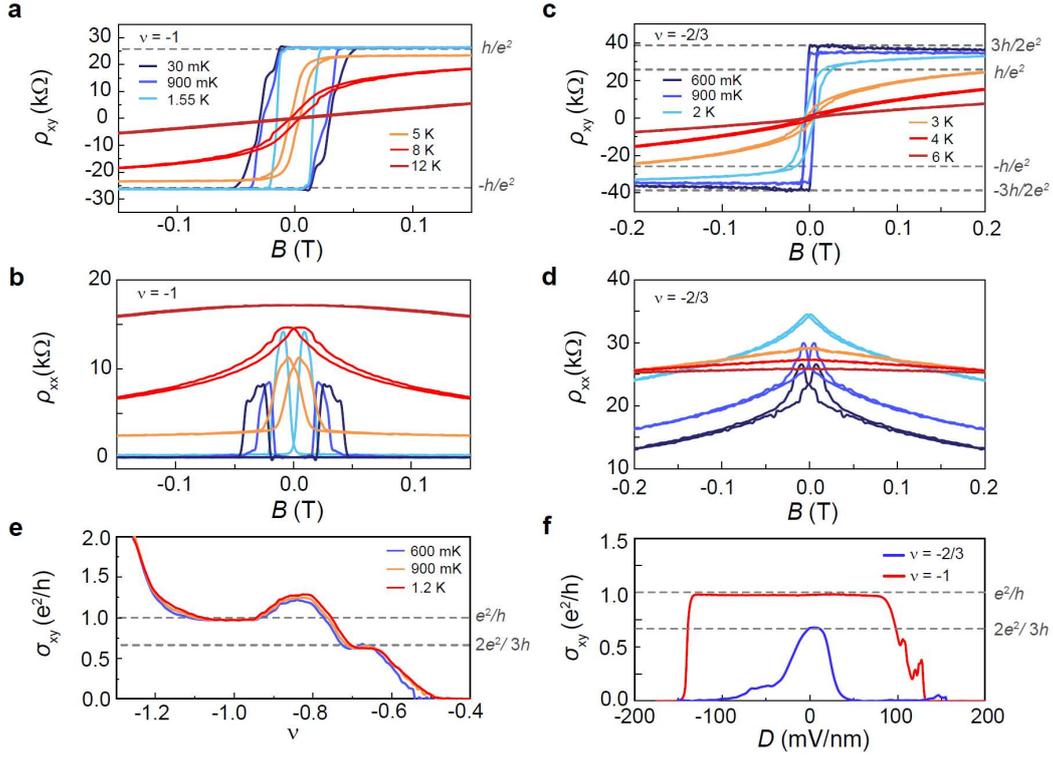

**Figure 2 | IQAH effect at $v = -1$ and FQAH effect at $v = -2/3$. a, b,** Magnetic-field dependence of $\rho_{xy}$ (**a**) and $\rho_{xx}$ (**b**) of the IQAH effect at $v = -1$ and $D = -115$ mV/nm at varying temperatures. Quantized $\rho_{xy}$ of $h/e^2$ and vanishing $\rho_{xx}$ are observed below 2 K at zero magnetic field. **c, d,** Magnetic-field dependence of $\rho_{xy}$ (**c**) and $\rho_{xx}$ (**d**) of the FQAH effect at $v = -2/3$ and $D = 5$ mV/nm at varying temperatures. Nearly quantized $\rho_{xy}$ of $3h/2e^2$ can be observed. **e,** Hall conductivity $\sigma_{xy}$ as a function of $v$ at $T = 600$ mK, 900 mK, and 1.2 K with $D = 5$ mV/nm. The $\sigma_{xy}$ is obtained using the antisymmetrized $\rho_{xy}$ and symmetrized $\rho_{xx}$ at $B = \pm 0.3$ T. Distinct $\sigma_{xy}$ plateaus can be observed around $v = -1$ and $-2/3$, with values closely approaching the quantized values of $e^2/h$ and $2e^2/3h$, respectively. **f,** $\sigma_{xy}$ as a function of $D$ measured at 900 mK at $v = -1$ and $v = -2/3$. The $\sigma_{xy}$ remains at the quantized values within certain $D$ range, then rapidly vanish as $D$ reach $D_c$. For $v = -1$, $\sigma_{xy}$ is derived from measured $\rho_{xy}$ and $\rho_{xx}$ at $B = 0$ T; and for $v = -2/3$, $\sigma_{xy}$ is obtained from antisymmetrized $\rho_{xy}$ and symmetrized $\rho_{xx}$ at $B = \pm 0.3$ T.

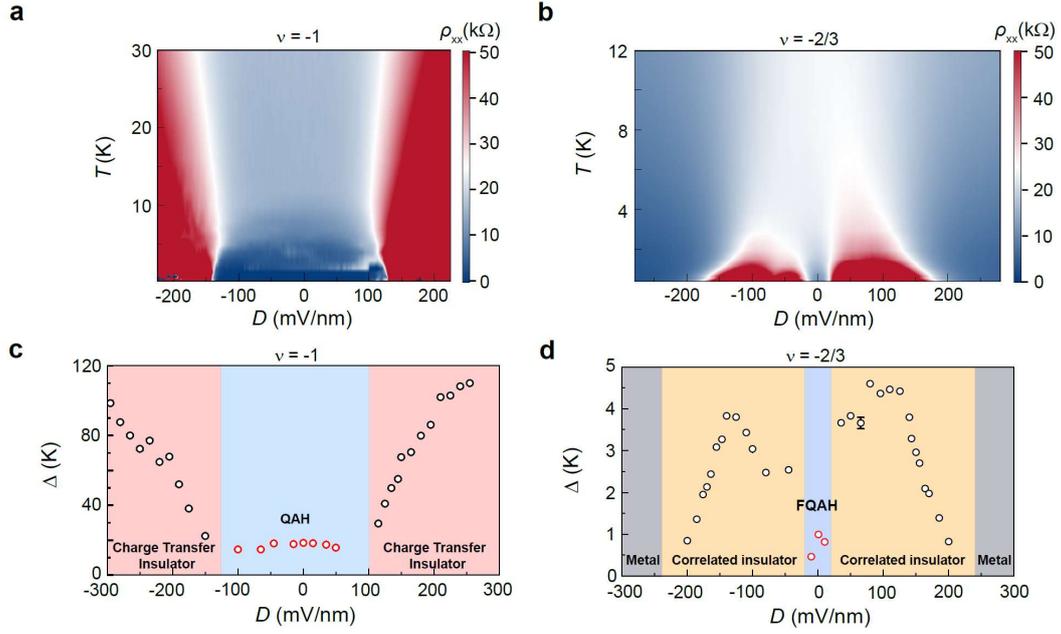

**Figure 3 | *D*-field tuned topological phase transitions. a, b,** $\rho_{xx}$ map as a function of *D* and *T* at $v = -1$ (**a**) and $v = -2/3$ (**b**). The $\rho_{xx}$ of $v = -1$ is measured at $B = 0$, and The $\rho_{xx}$ $v = -2/3$ is symmetrized at $B = \pm 0.3$ T. **c, d,** Extracted energy gap $\Delta$ (circles) by thermal activation fitting, as a function of *D*, at $v = -1$ (**c**) and $v = -2/3$ (**d**), respectively. Distinct phases are differentiated by various colors. The error bars smaller than the size of the circles are not displayed.

# Supplemental Figures for

# "Observation of integer and fractional quantum anomalous Hall effects in twisted bilayer MoTe$_2$"

Fan Xu[1,2], Zheng Sun[1], Tongtong Jia[1], Chang Liu[1], Cheng Xu[3,4], Chushan Li[1], Yu Gu[1], Kenji Watanabe[5], Takashi Taniguchi[6], Bingbing Tong[7], Jinfeng Jia[1,2,8], Zhiwen Shi[1,2], Shengwei Jiang[1,2], Yang Zhang[3,9]*, Xiaoxue Liu[1,2,8]*, Tingxin Li[1,2,8]*

[1]Key Laboratory of Artificial Structures and Quantum Control (Ministry of Education), Shenyang National Laboratory for Materials Science, School of Physics and Astronomy, Shanghai Jiao Tong University, Shanghai 200240, China.
[2]Tsung-Dao Lee Institute, Shanghai Jiao Tong University, Shanghai, 201210, China.
[3]Department of Physics and Astronomy, University of Tennessee, Knoxville, TN 37996, USA
[4]State Key Laboratory of Low Dimensional Quantum Physics and Department of Physics, Tsinghua University, 100084 Beijing, China
[5]Research Center for Electronic and Optical Materials, National Institute for Materials Science, 1-2 Namiki, Tsukuba 305-0044, Japan
[6]Research Center for Materials Nanoarchitectonics, National Institute for Materials Science, 1-1 Namiki, Tsukuba 305-0044, Japan
[7]Beijing National Laboratory for Condensed Matter Physics and Institute of Physics, Chinese Academy of Sciences, Beijing 100190, China
[8]Hefei National Laboratory, Hefei 230088, China
[9]Min H. Kao Department of Electrical Engineering and Computer Science, University of Tennessee, Knoxville, Tennessee 37996, USA

*Emails: txli89@sjtu.edu.cn; xxliu90@sjtu.edu.cn; yangzhang@utk.edu

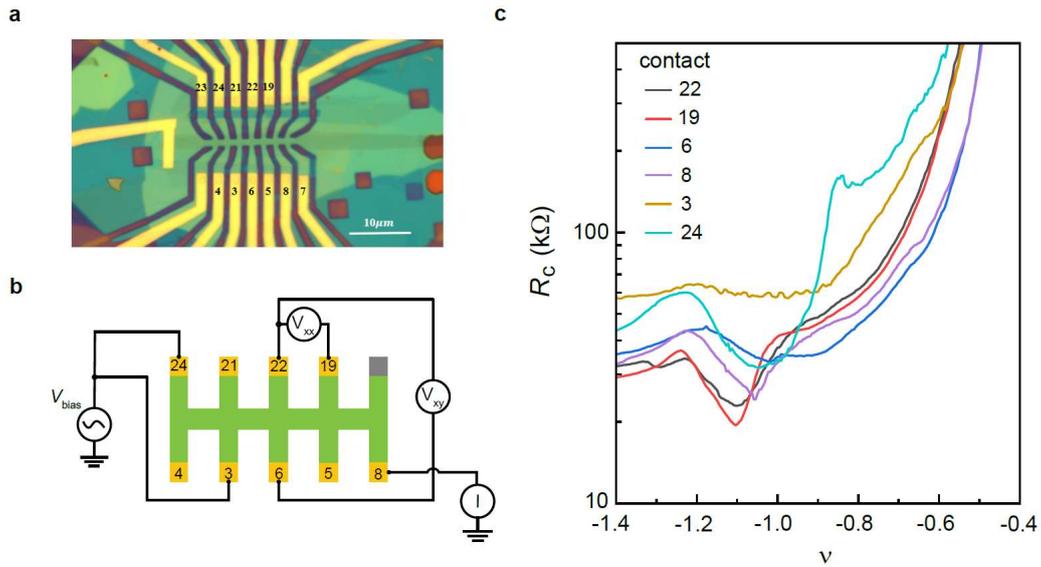

**Supplementary Figure 1 | Optical image and measurement configuration. a**, Optical micrograph of device I. The Hall bar geometry is defined by standard EBL and RIE processes. The scale bar is 10 μm. We found the middle region of the device (between contact 22,6 and 19, 5) is relatively uniform. **b**, Schematic figure of the measurement configuration. For most results presented in the main text, electrode 8 is grounded; electrode 24 and 3 are used as a source electrode. The longitudinal voltage drop is measured between 22 and 19; and the transverse voltage drop is measured between 22 and 6. **c**, Contact resistance versus filling factors at $T$ = 1.5 K and $D$ = 5 mV/nm. During the measurement of contact resistance, only the specific contact is biased with a 2 mV DC voltage while all other contacts are grounded. The DC current versus filling factors is measured using a source meter. At lower temperatures, the contact resistance at lower fillings increases more rapidly, going from ~ 100 kΩ at $v$ = -2/3 at 1.5 K to ~500 kΩ at $v$ = -2/3 at 600 mK.

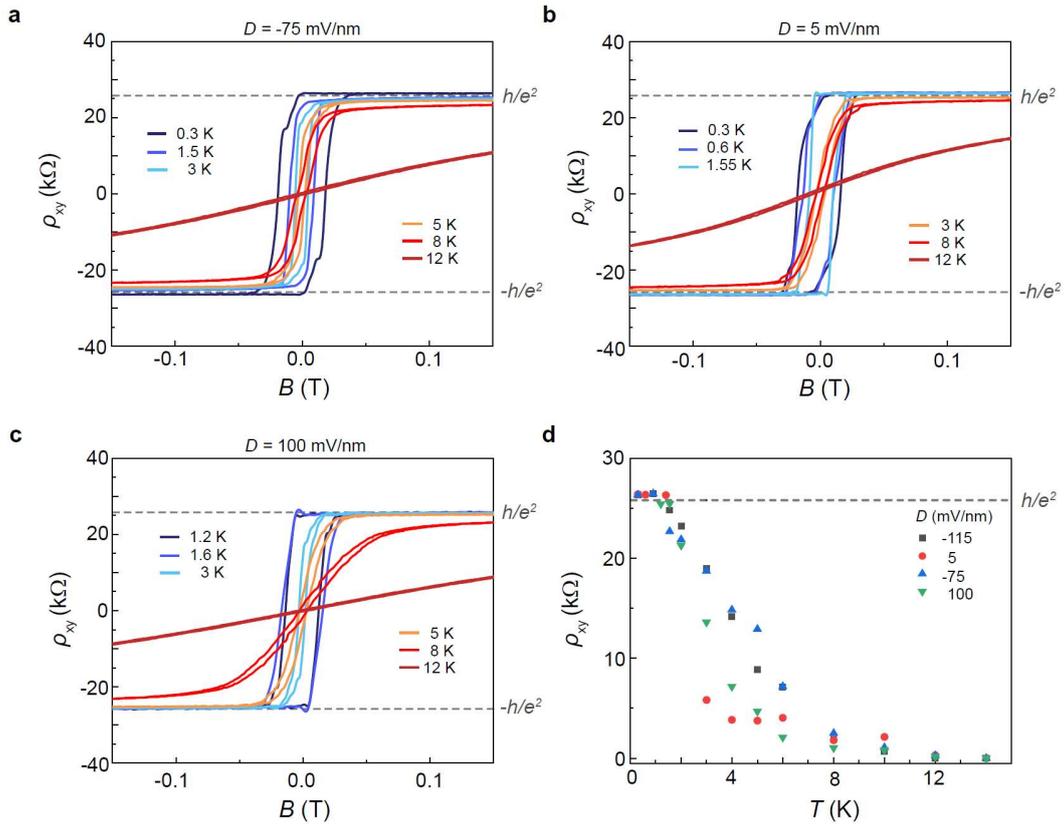

**Supplementary Figure 2 | IQAH effect under different *D*-field. a, b, c,** Magnetic-field dependence of $\rho_{xy}$ of the $\nu = -1$ IQAH effect at $D = -75$ mV/nm (**a**), 5 mV/nm (**b**), and 100 mV/nm (**c**) at varying temperatures. Quantized $\rho_{xy}$ of $h/e^2$ has been observed at all measured $D$ values at low temperatures. **d**, Temperature dependence of the zero-magnetic-field values of $\rho_{xy}$ of the $\nu = -1$ IQAH effect. The onset of magnetic ordering (Curie temperature) is at approximately 10–12 K and quantization of $\rho_{xy}$ happens below approximately 1-2 K, depending on specific $D$ values.

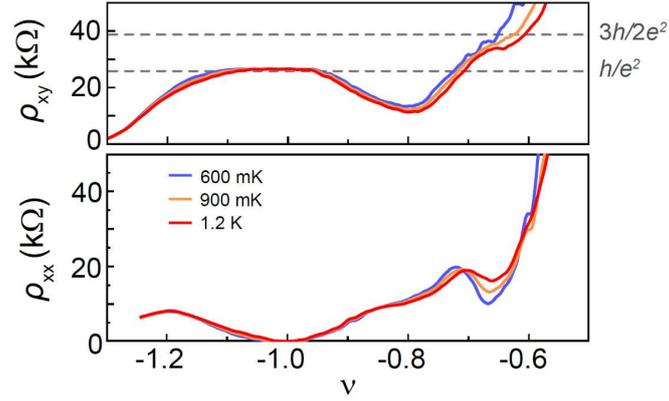

**Supplementary Figure 3 | Filling dependent Hall resistance and longitudinal resistance.** $\rho_{xy}$ (upper panel) and $\rho_{xx}$ (lower panel) as a function of $v$ at $T$ = 600 mK, 900 mK, and 1.2 K with $D$ = 5 mV/nm. The $\rho_{xy}$ ($\rho_{xx}$) is antisymmetrized (symmetrized) at $B = \pm\, 0.3$ T. $\rho_{xx}$ shows clear dips around $v = -1$ and $v = -2/3$ due to the chiral edge transport. $\rho_{xy}$ quantized at $h/e^2$ approximately from $v = -1.1$ to $-0.95$, and the quantized range become narrower with increasing temperature. Around $v = -2/3$, a narrower $\rho_{xy}$ plateau can be roughly identified above 900 mK, and the $\rho_{xy}$ value of the plateau is significantly larger than $h/e^2$ but still about 5%-15% smaller than the expected quantized value of $3h/2e^2$. Below 900 mK, the $\rho_{xy}$ plateau around $v = -2/3$ becomes indistinguishable, presumably due to the contact issue and sample inhomogeneities.

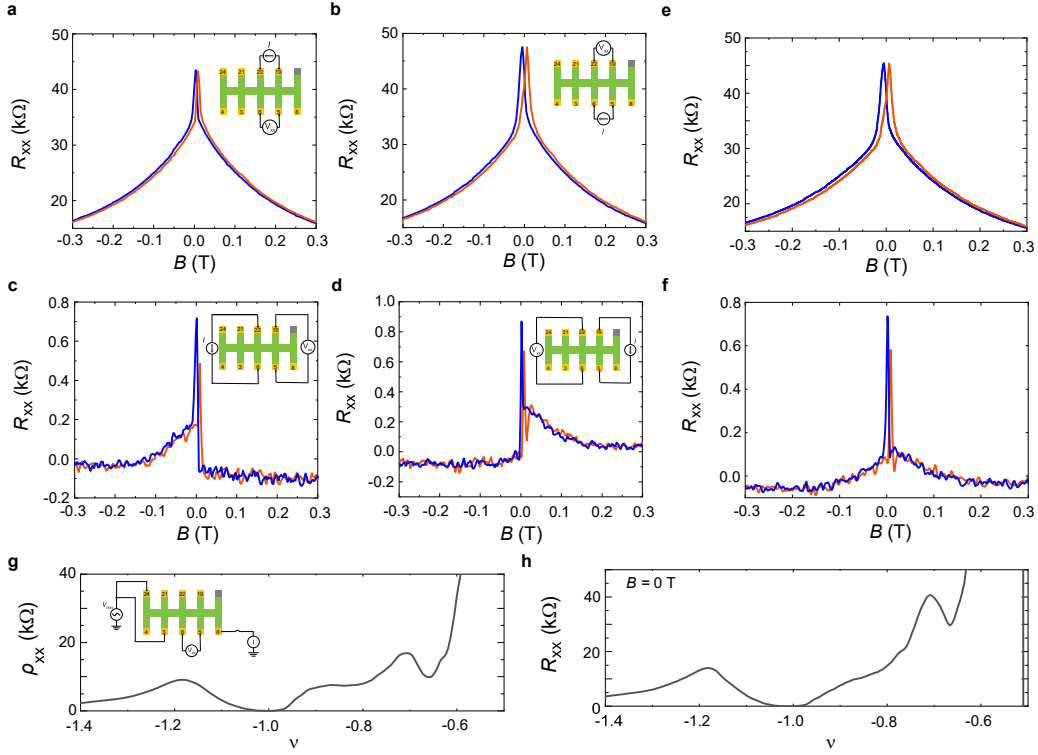

**Supplementary Figure 4 | Longitudinal resistance measured in different configurations. a-d,** Longitudinal resistance $R_{xx}$ measured with contacts 22, 19, 5, and 6 under different configurations at $\nu \sim -2/3$ at 600 mK. The specific current bias and voltage measurement configurations are indicated in the insets. **e,** Symmetrized $R_{xx}$ as a function of $B$ obtained from (**a**) and (**b**) using Onsager reciprocal relations $R_{xx} = \frac{R_{19-22,5-6} + R_{5-6,19-22}}{2}$. **f,** Symmetrized $R_{xx}$ obtained from (**c**) and (**d**) using Onsager reciprocal relations $R_{xx} = \frac{R_{22-6,19-5} + R_{19-5,22-6}}{2}$. **g,** Longitudinal sheet resistance $\rho_{xx}$ (symmetrized at $B = \pm 0.3$ T) as a function of $\nu$ measured at 600 mK with another pair of contacts, as illustrated in the inset. **h,** Onsager symmetrized $R_{xx} = \frac{R_{19-22,5-6} + R_{5-6,19-22}}{2}$ as a function of $\nu$ at $B = 0$ T and $T = 600$ mK. Distinct $R_{xx}$ minima can be identified around $\nu = -1$ and $-2/3$ in both (**g**) and (**h**).

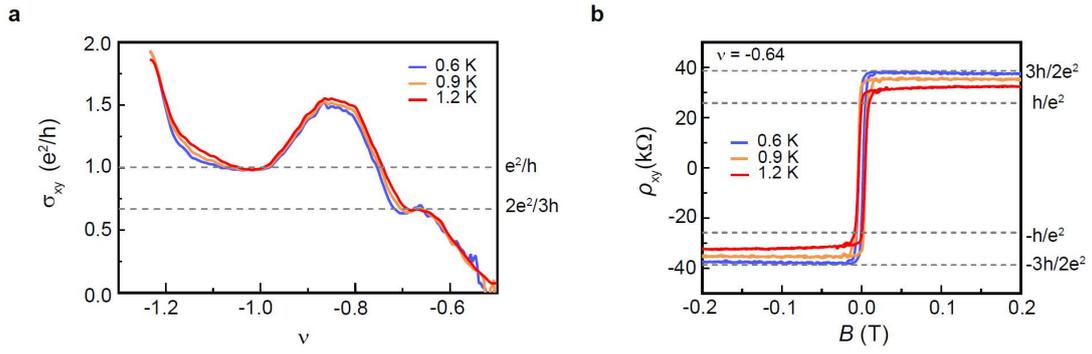

**Supplementary Figure 5 | Hall resistance measured with another pair of contacts. a,** Hall conductivity $\sigma_{xy}$ as a function of $v$ at $T = 600$ mK, 900 mK, and 1.2 K measured from contacts 19-5 with $D = 5$ mV/nm. The $\sigma_{xy}$ is obtained using the antisymmetrized $\rho_{xy}$ and symmetrized $\rho_{xx}$ at $B = \pm 0.3$ T. **b,** Magnetic-field dependence of $\rho_{xy}$ at $v = -0.64$ at varying temperatures measured from contacts 19-5 with $D = 5$ mV/nm. Overall, the results from contacts 19-5 are essentially the same as those from contact 22-6 (shown in main Fig. 2).

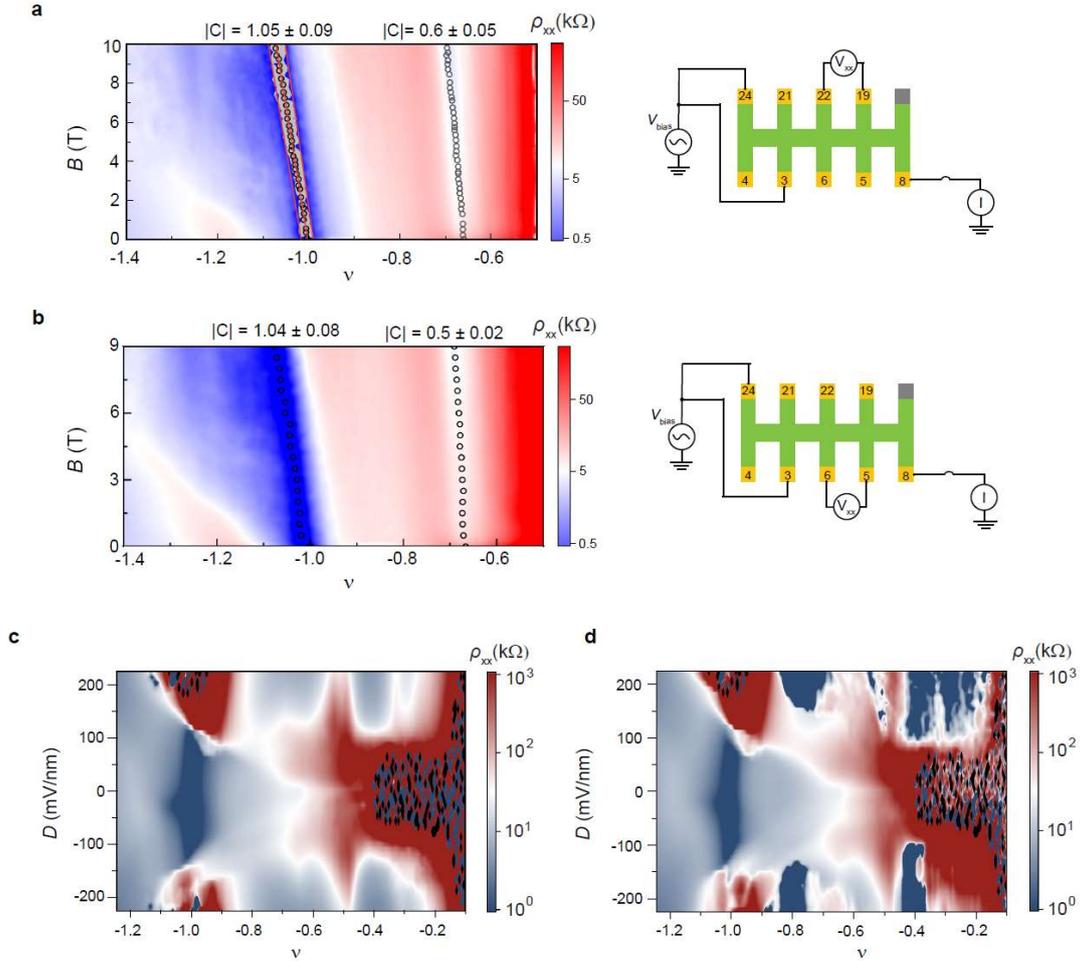

**Supplementary Figure 6 | Streda formula fitting of $\rho_{xx}$.** The $\rho_{xx}$ map as a function of $v$ and $B$ at 600 mK with different pairs of contacts. Empty circles mark the positions of $\rho_{xx}$ dips around $v = -1$ and $-2/3$. The extracted Chern numbers and errors are also displayed in the figures. The deduced Chern numbers show non-negligible deviations from the expected Chern number values. Several factors contribute to this deviation, including disorder effects and sample inhomogeneities. In addition, the broad nature of the $v = -1$ feature in $\rho_{xx}$, especially under magnetic fields, makes it challenging to accurately determine the actual position of the $v = -1$ gap. As depicted in (**a**), the grey region represents the region of vanishing $\rho_{xx}$ around $v = -1$, where the $\rho_{xx}$ value is smaller than 20 Ω. The three red lines around $v = -1$ in (**a**) represent possible unbiased fitting lines based on the grey region, yielding Chern numbers of 1.13, 1.09, and 0.97, from left to right, respectively. We therefore assigned the fitting values and errors by averaging possible fitting results. **c**, **d**, $\rho_{xx}$ (symmetrized at $B = \pm 0.1$ T) map as a function of $v$ and $D$ using (**c**) contacts 22-19 and (**d**) contacts 6-5 as voltage probes, measured at 1.5 K. The data from contacts 6-5 exhibit lower quality compared to contacts 22-19, which may be attributed to the characteristics of contact 5.

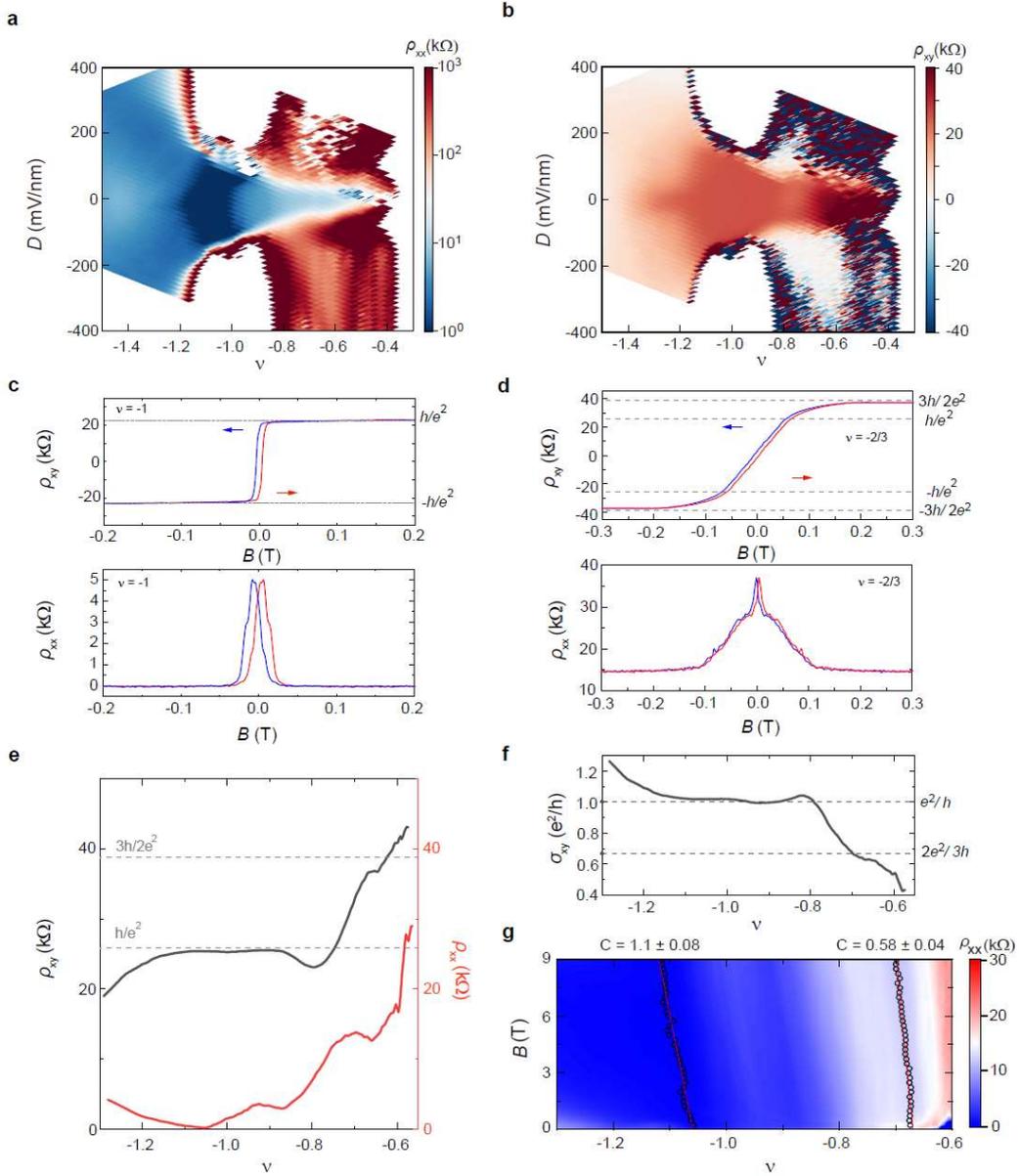

**Supplementary Figure 7 | Data of device II. a**, Symmetrized $\rho_{xx}$ and **b**, antisymmetrized $\rho_{xy}$ under $B = \pm 1$ T as a function of $v$ and $D$ at 1.5 K of device II. The twisted angle of this device is determined to be ~ 3.85 degrees. While the general patterns of the $\rho_{xx}$ and $\rho_{xy}$ map are consistent with device I, the finer details appear less distinct, likely due to larger inhomogeneities and disorder effects in device II. **c**, Magnetic-field dependence of $\rho_{xy}$ (upper panel) and $\rho_{xx}$ (lower panel) at $v = -1$ Quantized $\rho_{xy}$ of $h/e^2$ and vanishing $\rho_{xx}$ can be observed under a small magnetic field $B \sim 20$ mT. **d**, Magnetic-field dependence of $\rho_{xy}$ (upper panel) and $\rho_{xx}$ (lower panel) at $v = -2/3$. **e**, Symmetrized $\rho_{xx}$ and antisymmetrized $\rho_{xy}$ at $B = \pm 1$ T as a function of $v$. **f**, Filling dependent $\sigma_{xy}$ derived by symmetrized $\rho_{xx}$ and antisymmetrized $\rho_{xy}$ at $B = \pm 1$ T. **g**, Streda formula fitting of $\rho_{xx}$ of device II. (**c**)-(**g**) are measured at 1.5 K with $D \sim 0$ V/nm.

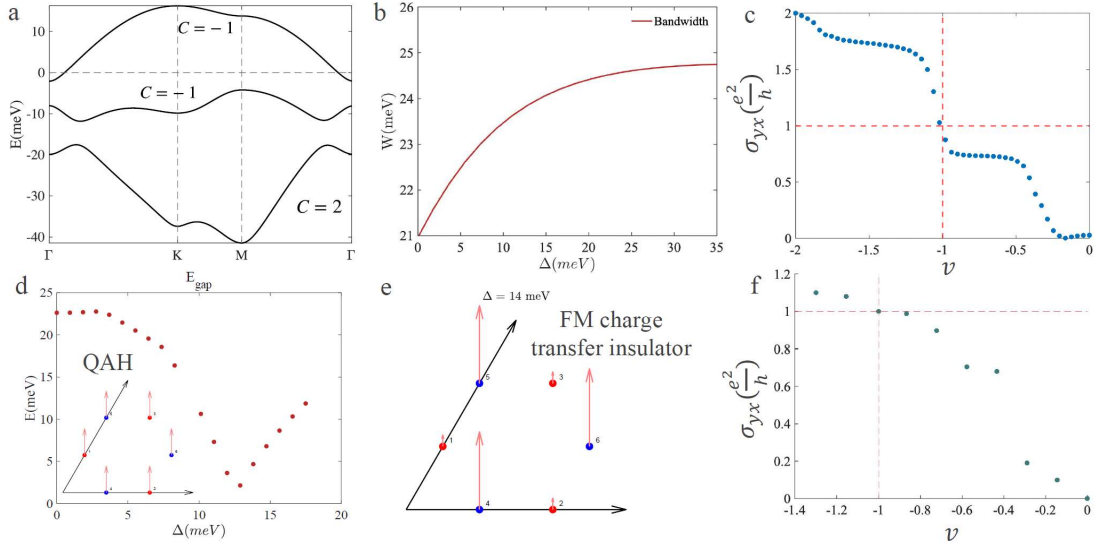

**Supplementary Figure 8 | Wannier tight-binding model and Hartree-Fock mean field simulation. a,** Wannier tight-binding band structure of K valley at twist angle $\theta = 3.7°$, which is identical with continuum model. The corresponding Chern number for three bands are -1, -1, 2. **b,** Bandwidth of top moiré valence band under increasing layer potential difference $\Delta$ (gating field). **c,** Anomalous Hall conductance (Berry curvature integral) vs. filling factor $\nu$ from non-interacting band structure of K valley. We note that Berry curvature distribution is non-uniform across the Brillouin zone. **d,** Topological transition under gating field, from $C_{2y}$ symmetric quantum anomalous Hall state to ferromagnetic charge transfer insulator. The insulating gap is calculated from the Hartree-Fock method. We note that the transferred charges increase with the gating field, and this is a continuous second-order transition. The insulating gap of the charge transfer insulator is proportional to the gating field. **e,** Charge and spin distribution of ferromagnetic charge transfer insulator near transition field. **f,** Anomalous Hall conductance vs. filling factor $\nu$ from Hartree-Fock simulation of 2x2 unit cell. We note that the Hartree-Fock captures the essential behavior of integer quantum anomalous Hall and the anomalous Hall metal regime.

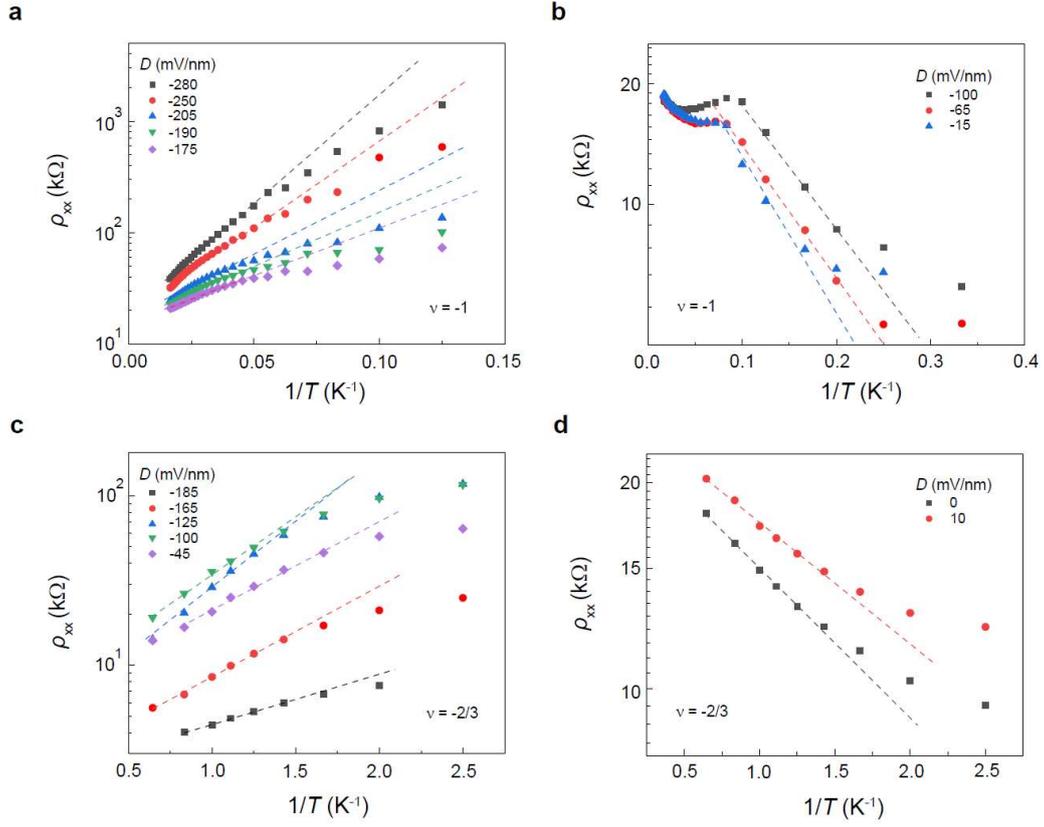

**Supplementary Figure 9 | Estimation of charge gaps. a, b,** Arrhenius plots of $\rho_{xx}$ at varying electric field at $v = -1$, for the charge transfer insulator (**a**) and IQAH state (**b**), respectively. Dashed lines are the fit to the thermal activation behavior $\rho_{xx} \sim e^{-2\Delta/k_B T}$, with $\Delta$ and $k_B$ denoting the charge gap and Boltzmann constant, respectively. **c, d,** Same as **a, b**, but for the $v = -2/3$ correlated trivial insulator (**c**) and FQAH state (**d**). The extracted charge gap values are shown in Fig. 3**c** and 3**d** of the main text.